# Direct observation of patterned self-assembled monolayers and bilayers on silica-on-silicon surfaces

Hadas Alon,[1,2] Idan Bakish,[1] Josh Nehrer,[1] Assaf Y. Anderson,[2] Chaim N. Sukenik,[2] Avi Zadok,[1] and Doron Naveh[1,*]

[1]*Faculty of Engineering and Institute for Nanotechnology and Advanced Materials, Bar-Ilan University, Ramat-Gan 5290002, Israel*
[2]*Department of Chemistry and Institute for Nanotechnology and Advanced Materials, Bar-Ilan University, Ramat-Gan 5290002, Israel*
[*]*Doron.Naveh@biu.ac.il*

**Abstract:** Self-assembled monolayers (SAMs) of organic molecules are widely employed in surface chemistry and biology, and serve as ultra-fine lithographic resists. Due to their small thickness of only a few nanometers, the analysis of patterned monolayer surfaces using conventional methods requires thorough point-by-point scanning using complicated equipment. In the work reported herein, patterned monolayers are simply and directly observed using a bright-field optical microscope. The monolayers modify the spectral reflectivity pattern of a silica-on-silicon thin film, and introduce a contrast between bare and monolayer-coated regions of the substrate. The method can also distinguish between regions of single-layer and bi-layer coatings. The observations are supported by calculations, and by control experiments using atomic force microscopy, scanning Raman spectrometry and scanning reflection spectrometry. The results are useful for electro-optic devices, selective wafer-bonding protocols and lab-on-a-chip test systems. We show here that chemical reactions leading to the formation of a bi-layer of SAMs correspond to an optical contrast visible to the naked eye, enabling such detection to provide a simple, yet effective differentiation between monolayers and adsorbed analytes with possible applications for chemical and/or biological sensing.

## 1. Introduction

Self-assembled monolayers (SAMs) are ordered, single-molecular layers of organic materials which may form spontaneously on various surfaces, in solution or in the gas phase [1,2]. These thin films have proven to be powerful tools for controlling surface chemistry and are the basis of applications ranging from sensors [3-5] to controlling surface free energy and adhesion. The patterning of SAMs is of interest for diverse potential applications such as ultra-thin lithographic resists [6], positioning and attachment of different particles [7] and electronic molecular devices [8].

Many methods are routinely used for characterizing the thickness, composition and order of SAMs, such as ellipsometry [9], atomic force microscopy (AFM) [10], Fourier-transform infra-red spectroscopy (FTIR) [11], x-ray photo-electron spectroscopy (XPS) [12], and scanning tunneling microscopy (STM) [13]. Contact angle goniometry measurements are widely used in the estimation of surface free energy of the monolayers [14]. Some of these methods are inherently statistical, and are based on an averaging over large uniform surfaces, whereas others are extremely localized and require a thorough, nanoscale point-by-point sampling. In addition, these observation methods require sophisticated and expensive equipment. Some of them involve meticulous sample preparation, and/or could be destructive to the sample being examined. These characterization methods do not provide a rapid, convenient analysis of patterned SAM-coated surfaces.

Spectral analysis of reflections from one or a few molecular layers can resolve changes in optical path length on a sub-nanometer scale [15-19]. Gauglitz and coworkers employed spectral interferometry to monitor the swelling of thin polymer films exposed to different analytes, as well as antigen-antibody reactions [15]. A review of chemical and biological applications of spectral interferometry point-sensors is provided in [16]. Reflective interferometry was extended to the spatially resolved analysis of SAM patterns on a substrate, in a significant series of works by Rothberg and associates [17-19]. The deposition of molecular films could be observed with sub-Angstrom-level resolution based on relatively large contrast [17-19]. These experiments required polarization control, careful collimation and angular alignment of the interrogating beam, and narrow-band optical filtering. The lateral resolution of the measurements was limited to a length scale on the order of tens of microns. The direct imaging of sub-micron SAM patterns using a simple, readily available setup has not yet been reported.

Similar methodology has been applied to the characterization of exfoliated graphene films [20], in which small regions of single-atomic layers must be identified. Over the last few years, several groups reported the direct observation of single-layer graphene using standard, bright-field microscopy, when the films are deposited on silica-on-silicon substrates [21-23]. The thickness of the silica layer is designed to provide wavelength-varying reflectivity within the visible range. The presence of a graphene monolayer slightly modifies the spectral reflectivity pattern. Although small, these spectral variations can be recognized by the naked eye [20,24]. The technique can also distinguish between regions containing different numbers of graphene layers [20,24]. The direct observation of nanometric coatings of lossy materials using visible light was recently reported as well [25]. A similar technique was used by Daaboul *et al.* in the sizing of micro-spheres with 70-200 nm diameters on a silica-on-silicon substrate [26], and in the detection and classification of viruses of a similar size [27].

In this work, we employ reflective interferometry in the analysis of patterned SAMs on a silica-on-silicon substrate .500 nm-wide features are recognized in a standard bright-field microscope image, operating at normal incidence without polarization control or spectral filtering. The interferometry contrast is sufficient to resolve single layers and bi-layers of SAM-forming molecules using a simple CMOS imaging sensor. The direct observation is compared with, and validated by, AFM analysis, Raman scattering microscopy and position-dependent reflection spectrometry. These observations are unequivocally related to the presence of a monolayer and can, with simple visible light detection, 'see' organic layers that are only a few nanometers thick.

## 2. Theory

Reflection from a SAM deposited on silica-on-silicon is illustrated in Fig. 1. The monolayer and the sub-micron silica layer are treated as non-absorbing, dielectric materials. Normal incidence is assumed throughout this work, for convenience and simplicity.

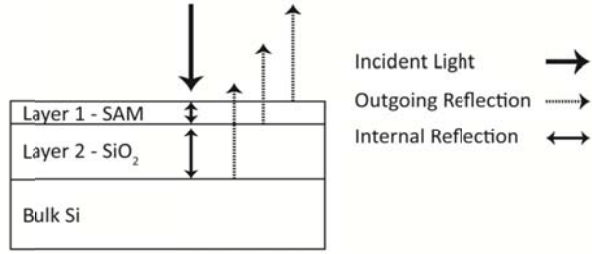

Fig. 1. Illustration of multiple reflections from thin layers of SAM and SiO$_2$ on a silicon substrate

The reflectivity of the composite structure can be expressed in terms of cascaded elementary transmission matrices of individual thin films. The transverse electric and magnetic fields reflected from the thin film assembly are given by [28,29]:

$$\begin{bmatrix} B \\ C \end{bmatrix} = \begin{bmatrix} \cos\delta_1 & \dfrac{i\sin\delta_1}{\eta_1} \\ i\eta_1 \sin\delta_1 & \cos\delta_1 \end{bmatrix} \begin{bmatrix} \cos\delta_2 & \dfrac{i\sin\delta_2}{\eta_2} \\ i\eta_2 \sin\delta_2 & \cos\delta_2 \end{bmatrix} \begin{bmatrix} 1 \\ \eta_3 \end{bmatrix} \quad (1)$$

Here $B$ and $C$ denote the electric and magnetic fields at the interface between air and the upper SAM layer. $\delta_{1,2} = 2\pi n_{1,2} d_{1,2} / \lambda$ are the optical phases acquired in the passing of light through the SAM and the SiO$_2$ layer, respectively, $n_{1,2}$ denote the refractive indices of the two media, $d_{1,2}$ indicate their thicknesses, and $\lambda$ is the vacuum wavelength of the incident light. $\eta_{1,2,3} = n_{1,2,3} \eta_0$ denote the optical admittances of the monolayer, silica layer, and bulk silicon, respectively, defined as the ratio between the magnitudes of magnetic and electric fields components. The admittance of vacuum is noted by $\eta_0$. The admittance of the composite thin film assembly is given by [28,29]:

$$Y = \frac{C}{B} \quad (2)$$

The overall power reflectivity coefficient of the three-layer stack can be expressed as:

$$R = |r|^2 = \left|\frac{\eta_0 - Y}{\eta_0 + Y}\right|^2 \quad (3)$$

Direct observation of SAM patterns relies on the contrast in reflectivity between regions where a monolayer is deposited, and regions which are monolayer-free:

$$Contrast = \frac{R(\text{Air}) - R(\text{SAM})}{R(Air)} \quad (4)$$

Figure 2 shows the calculated contrast map, as a function of the silica layer thickness $d_2$ and the incident wavelength $\lambda$. The refractive indices of silica and silicon as a function of wavelength were taken from known references [30,31]. The refractive index of a molecularly thin organic monolayer created using octadecyltrichlorosilane (OTS) was estimated as $n_1 = 1.47$ using the group contribution method [32]. The thickness of an OTS monolayer is typically measured by ellipsometry to be around 2.5 nm [33]. The calculated contrast reaches ±7%. For each thickness of the silica layer, the contrast reaches a maximum at different wavelengths.

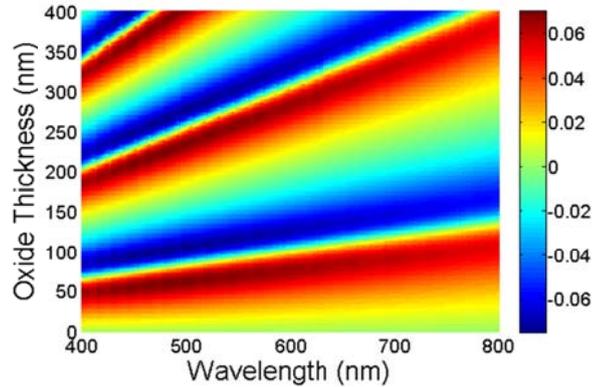

Fig. 2. Calculated maps of the expected contrast between the reflectivity of a bare silica-on-silicon sample, and a sample that is coated with a 2.5 nm-thick SAM created using OTS. Contrast is given as function of the silica layer thickness and the incident wavelength.

### 3. Results

The monolayers used in our work were fabricated using OTS and methyl 11-(trichlorosilyl)undecanoate. OTS is known to readily form dense and well-ordered monolayers [34]. The methyl 11-(trichlorosilyl)undecanoate SAMs are the basis for a simple method for bi-layer assembly whose basic principles were first demonstrated by Ulman and coworkers [35] and that we have adapted for use with a shorter, more readily available SAM forming material. Details of monolayer synthesis and deposition are provided in the Appendix.

Monolayers were patterned through lithography and lift-off processes. Both point-by-point electron-beam lithography and photo-lithography through pre-patterned masks were employed (details of the lithography are provided in the Appendix). Monolayers were deposited on a silica-on-silicon substrate, with a silica layer thickness of 298±6 nm as measured by ellipsometry.

Figure 3 shows an optical microscope image of an OTS-coated substrate, taken with an objective lens of 100X magnification. The SAM consists of multiple features, between 200 nm and 10 μm in size, which were patterned using electron-beam lithography. Patterns 500 nm or wider are clearly recognized by direct observation through a bright-field microscope. Figure 4 shows AFM scans of the finer patterns of Fig. 3, having widths of 200 nm and 500 nm. The profile maps obtained by the AFM match the designed patterns as well as the direct microscope observations. The height of the patterns is approximately 2.5 nm, in good agreement with the expected height of an OTS monolayer [33].

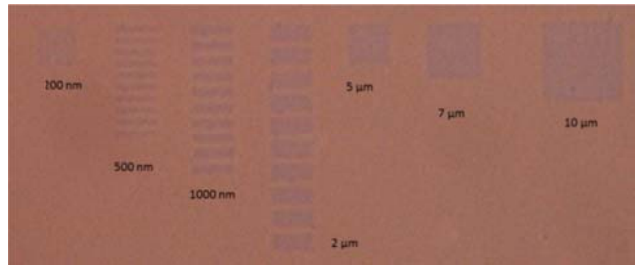

Fig. 3. Bright-field microscope image of multiple patterns, defined in an OTS monolayer through electron-beam lithography. Scales are noted in the image. The magnification of the microscope was 100X.

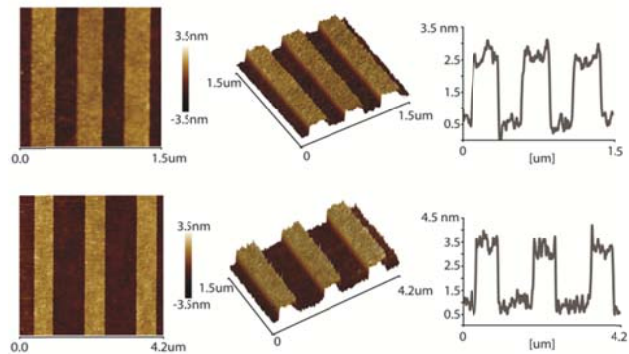

Fig. 4. AFM images and height scans of multiple patterns, defined in an OTS monolayer through electron-beam lithography. Top: feature width of 200 nm. Bottom: feature width of 500nm.

The observation of patterned monolayers was further demonstrated using 2.8 X 2.8 mm$^2$ diamond-shaped features in an OTS monolayer, patterned by photo-lithography. The patterns were characterized using direct observation, and also by position-dependent Raman spectroscopy and position-dependent visible light reflection spectroscopy. Figure 5 shows a clearly discernible direct optical image of the monolayer pattern boundary.

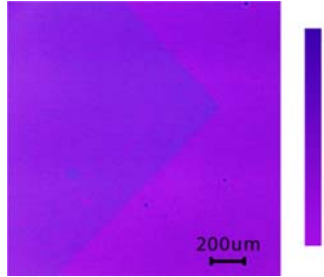

Fig. 5. Bright-field microscope image of the boundary between OTS SAM-coated and uncoated regions in a sample patterned via photo-lithography.

Figure 6 (a) shows an example of the Raman scattering spectrum, collected within the region that was covered by an OTS monolayer. The line at 2847 cm$^{-1}$ is characteristic of the symmetric CH$_2$ stretch in OTS [36]. A two-dimensional map of the strength of the 2847 cm$^{-1}$ peak, as a function of position across the sample, is shown in Fig. 6(b). The expected diamond-shaped pattern, directly observed in Fig. 5, also appears in the positional Raman scan. This result provides further indication that the observed pattern is indeed due to the presence of the OTS SAM.

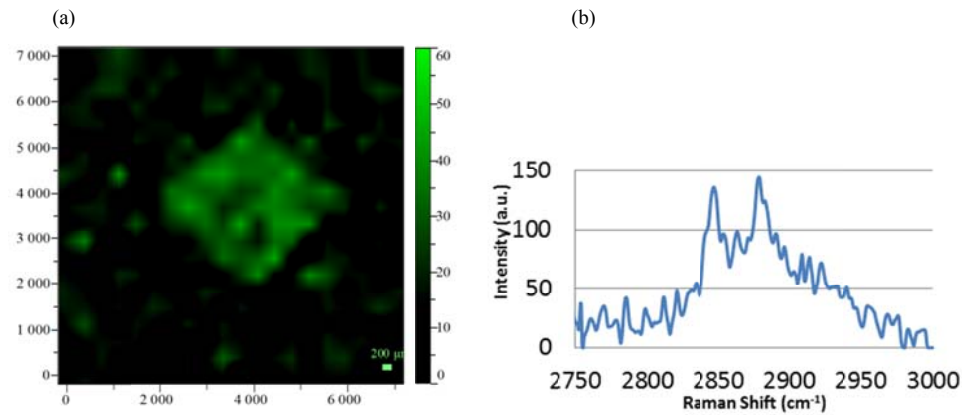

Fig. 6. (a) – Raman scattering spectrum collected from an OTS-coated region of a silica-on-silicon sample. A characteristic peak at 2847 cm$^{-1}$ is evident. (b) – map of the strength of the 2847 cm$^{-1}$ Raman peak as function of position across the sample. The expected diamond-shaped pattern, defined by photo-lithography, is apparent.

Control experiments that either left out the OTS from the deposition solution, or that replaced the silica-on-silicon sample by a bulk silicon sample with a 2 nm-thin native oxide layer, did not reveal any spatial patterns. This further strengthened the assertion that the observed patterns were due to the presence of the monolayer, and the significance of the multiple reflections in the layer stack as discussed above.

Next, the spectral reflectivity of the patterned sample was measured as a function of position, using a specialized scanning spectral reflectometer [37]. The spatial resolution of the measurements was 1 mm. Figure 7 shows the contrast between measured reflectivities of locations within (SAM-covered) and outside of (bare regions) the diamond-shaped pattern, as a function of wavelength. The corresponding, calculated spectral contrast is shown as well. A qualitative agreement between measurement and simulation is obtained, although the experimentally observed contrast is larger than expected. Figure 8 shows two-dimensional spatial maps of the reflectivity at 640 nm and 540 nm wavelengths, which correspond to maximum and minimum reflectivity of the monolayer coated region, respectively.

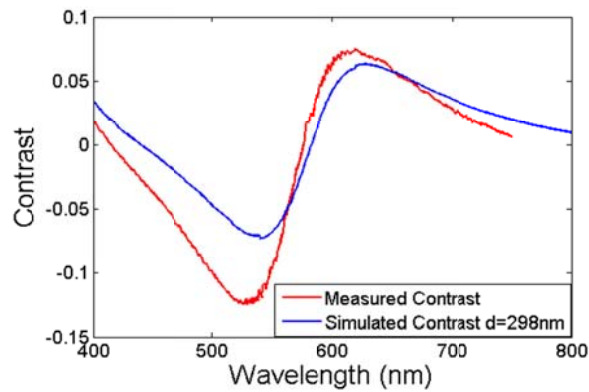

Fig. 7. Measured (red) and simulated (blue) contrast between the reflectivities of an OTS SAM-coated region and an uncoated region of a silica-on-silicon sample, as a function of incident wavelength. Calculations were performed for a silica layer thickness of 298 nm.

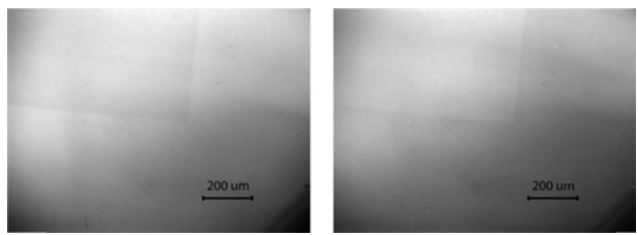

Fig. 8. Microscope images of a silica-on-silicon sample, filtered at 640 nm (left) and 540 nm (right) wavelengths. A diamond-shaped region at the center of the sample was covered by an OTS SAM. The edge of the region can be observed in the images. The coated region is characterized by a weaker reflectivity than that of its surroundings at 640 nm, and by a stronger reflectivity at 540 nm.

In order to investigate the contrast variations with thickness change, we fabricated two partially overlapping square patterns using a well-defined bilayer. An illustration describing the geometrical organization of the bilayer is given in Fig. 9. A bilayer is only formed in the region in which the two patterns overlap. Figure 10(a) shows an optical image, using a regular CMOS camera, of a patterned sample containing regions of a single and double layers of methyl 11-(trichlorosilyl)undecanoate, following the schematic in Fig. 9. Figure 10(b) shows a magnified image of a region containing patterns with zero, one or two layers; the magnified region is located on the upper right-hand corner in Fig. 10(a). An optical filter with a bandwidth of 10 nm and a central transmission wavelength of 570 nm was applied to the magnified image only. The transmission wavelength was chosen to match the maximum contrast between monolayer-coated and bare regions. Regions of a single layer and a bi-layer are clearly observed.

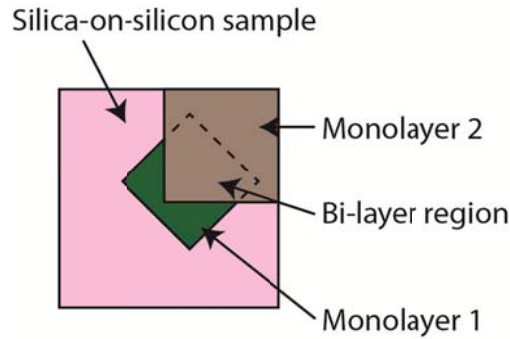

Fig. 9. Illustration of the bilayer deposition layout. The area of overlap between monolayers is the region in which the bi-layer is formed

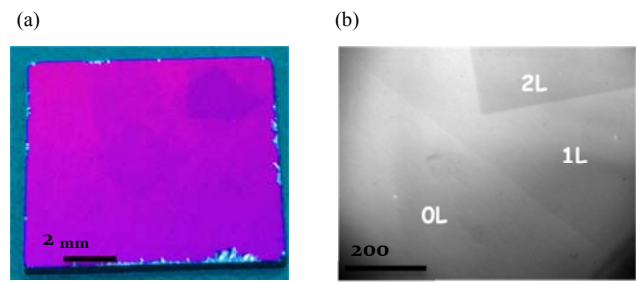

Fig. 10. (a) - a camera image of a silica-on-silicon sample, containing patterned regions of self-assembled single-layers and double-layers of methyl 11-(trichlorosilyl)undecanoate. (b) - enlarged image of the upper right-hand corner of the left-hand panel. A 10 nm-wide optical bandpass filter, centered at 570 nm, was used to enhance the contrast between regions.

## 4. Conclusions

The direct observation of 2.5 nm-thick, single molecular layers of organic materials, using standard optical microscopy, has been convincingly demonstrated. The observation principle was inspired by similar work analyzing graphene layers [20,24]. The monolayer deposition modifies the reflectivity spectrum of a silica-on-silicon substrate, providing distinguishable contrast between the exposed regions of the substrate and domains coated with various coatings. Both electron-beam lithography and optical lithography were used to make patterns of different sizes and shapes, and the observation of patterns as narrow as 500 nm was successfully demonstrated. Direct observation could also discern between regions of a single layer and of a bi-layer. The experimental observation is supported by numerical simulations, and corroborated by AFM analysis, scanning spectral reflection measurements, and scanning Raman scattering microscopy. Compared with previous reflective interferometry analyses of monolayers [17-19], the proposed setup is considerably simpler and provides order-of-magnitude higher spatial resolution, with compromised contrast and thickness sensitivity.

    The proposed observation method could be very useful for applications of patterned monolayers. The observed contrast between single and bi-layer regions paves the way for implementing a simple chemical sensor exhibiting a color change upon formation of a second layer on a monolayer interface. Such patterns of varying reflectivity would be visible to the naked eye and could provide simple, real time and high-throughput biological and chemical sensors [16,27].

**Appendix: experimental procedures**

Substrate preparation: Silica-on-silicon wafers with a silica layer thickness of about 300 nm were sonicated and scrubbed with acetone under cleanroom conditions. The wafers were then exposed to oxygen plasma cleaning for five minutes. Silicon wafers with native silica layer were cleaned in hexane, acetone and ethanol, blown dry in a filtered nitrogen stream and placed in a fresh piranha solution ($H_2SO_4$ :$H_2O_2$, with 70:30 volume ratio at 80° C) for 20 minutes. The wafers were rinsed 3 times with deionized, doubly-distilled water and dried with nitrogen.

Electron-beam lithography: Silica-on-silicon wafer dies of 1 X 1 $cm^2$ area were covered with a 250 nm-thick PMMA K950 A4 resist. Patterns were defined in the resist using a Crestec CABL-9500C electron-beam lithography machine. The lithography was performed at 50 KV with 1 nA current, and at 100 dots/μm resolution. Following the electron-beam writing, the samples were put in MIBK developer for 60 seconds and in isopropyl alcohol for 30 seconds, and then placed in a plasma chamber with 125 sccm flow of argon for duration of 30 seconds and at RF power of 100 Watts, for the cleaning of resist residues.

Photo-lithography: Silica-on-silicon wafer dies of 1 X 1 $cm^2$ area were spin-coated with Shipley 1813 photo-resist, baked for 2 minutes at 115 °C and exposed to UV light through a pre-patterned photo-lithographic mask. Illuminated areas of the photo-resist were developed in MF319 and doubly-distilled water, exposing the underlying silica layer.

Monolayer deposition and characterization: 10 mL of DCH and either 50 μL of OTS or 20 μL of methyl 11-(trichlorosilyl)undecanoate were placed in a clean, dry test tube under positive pressure of nitrogen, together with a silicon or silica-on-silicon sample. The samples were immersed in the solution for 75 minutes at room temperature, sonicated in chloroform for 6 minutes, placed in hexane for 6 min at 80° C, scrubbed with hexane at room temperature, and blown dry under a nitrogen stream. Silica-on-silicon samples which were previously covered by a lithographic resist were washed in acetone as well, to perform lift-off.

An OTS SAM on a silica-on-silicon substrate is illustrated in Fig. 11. The oxide thickness was measured using an Accurion Nanofilm ep3 ellipsometer. Raman Spectra were measured using a Raman microscope (HORIBA Scientific LabRAM HR). The shifts were measured from a laser with an excitation wavelength of 532 nm. Contact angle goniometry was used to verify the surface wettability and AFM was used for verification of the monolayer height after fabrication of patterned SAM.

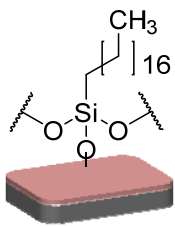

Fig. 11. Illustration of octadecyltrichlorosilane (OTS) monolayer on silica on silicon wafer

Bilayer formation (modeled after [35]): 10 mL of lithium aluminum hydride ($LiAlH_4$) solution (1 M in THF) were placed in a dry, clean test tube under a nitrogen stream. A methyl 11-(trichlorosilyl)undecanoate-modified sample was placed in the test tube for 10 minutes. Next, the sample was removed and placed in a 10% solution of hydrochloric acid for 5 minutes. The sample was then rinsed with hexane at room temperature and dried under a filtered nitrogen stream. A second layer of methyl 11-(trichlorosilyl)undecanoate was then deposited as illustrated in Fig. 12. FTIR traces of a silicon wafer, taken at different stages of a two-layer deposition process, are shown in Fig. 13.

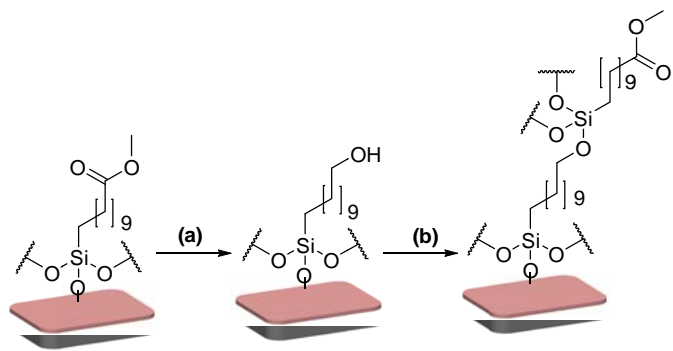

Fig. 12. Illustration of methyl 11-(trichlorosilyl)undecanoate bilayer formation on a silica-on-silicon sample. (a) denotes LiAlH$_4$, (b) denotes second layer of methyl 11-(trichlorosilyl)undecanoate.

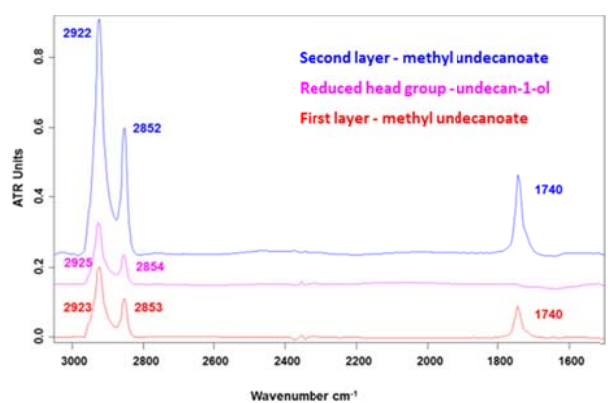

Fig. 13. FTIR monitoring of the construction of a bi-layer film. The appearance and disappearance of the peak at ~1740 cm$^{-1}$ indicates the deposition of the methyl 11-(trichlorosilyl)undecanoate and its subsequent reduction.

Synthesis of methyl undec-10-enoate [38]: Undec-10-enoic acid (7 g, 37.9 mmole), methanol (70 mL) and 2-3 drops of H$_2$SO$_4$ were placed into a 250 mL flask equipped with a magnetic stirring bar. The reaction mixture was stirred under reflux for 1-2 hours until the reaction was complete (monitored by TLC: silica gel/ hexane: ethyl acetate 2:1). The crude product was isolated by evaporating the excess methanol using a rotary evaporator. No further purification was needed before hydrosylilation. (yield: 7.53 g, 100 %). $^1$H-NMR: 5.8 (m, 1H, CH$_2$=CH), 4.7 (m, 2H, CH$_2$=CH), 3.66 (s, 3H, OCH$_3$), 2.3 (t, J = 8 Hz, 2H, CH$_2$C=O), 2.05 (m, 2H, CH$_2$=CH-CH$_2$), 1.6 (m, 2H, CH$_2$CH$_2$C=O), 1.21-1.4 (m, 10H). $^{13}$C-NMR: 25.084 (CH$_2$CH$_2$C=O), 29.01, 29.273 (2C), 29.352 (2C), 33.24 (CH$_2$C=O), 34.235 (CH$_2$=CHCH$_2$), 51.535 (OCH$_3$), 114.2 (CH$_2$=CH), 139.102 (CH$_2$=CH), 175.1 (C=O)

Synthesis of methyl 11-(trichlorosilyl)undecanoate [38]: Methyl undec-10-enoate (2 g, 10.08 mmole), HSiCl$_3$ (6 mL), and 2 drops of 4% solution of H$_2$PtCl$_6$*6H$_2$O in dry i-PrOH were placed into a 20 mL oven-dried pressure tube containing a magnetic stirring bar. The contents of the tube were stirred at room temperature overnight. The progress of the reaction was monitored by the disappearance of the olefinic protons in the $^1$H-NMR. After the reaction was complete, the contents of the tube were transferred (under nitrogen) to a 25 mL oven-dried round bottom flask. Excess HSiCl$_3$ was evaporated using a stream of nitrogen and the product was isolated by Kuglrohr distillation at 130 °C and 0.05 mm Hg pressure (yield: 2.35 g, 70 %). The synthesis is illustrated in Fig. 14. $^1$H-NMR: 3.66 (s, 3H, OCH$_3$), 2.3 (t, J = 8 Hz, 2H, CH$_2$C=O), 1.6 (m, 4H, CH$_2$CH$_2$C=O and Cl$_3$SiCH$_2$), 1.21-1.4 (m, 14H, aliphatic). $^{13}$C-NMR: 22.40, 24.522, 25.132, 29.192, 29.352 (2C), 29.481, 29.558 (2C), 34.235 (CH$_2$C=O), 51.535 (OCH$_3$), 175.102 (C=O).

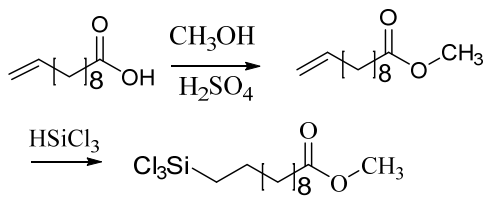

Fig. 14. Synthesis of methyl 11-(trichlorosilyl)undecanoate

**Acknowledgments**

I. Bakish and A. Zadok acknowledge the financial support of the Israeli Science Foundation (ISF) under grant 635/10, and of the Chief Scientist Office of the Israeli Ministry of Economy within the 'TERA SANTA' consortium. H. Alon and C. N. Sukenik acknowledge the support of the ISF under grant 1826/12, and of the Edward and Judith Steinberg Chair in Nano-Technology, Bar-Ilan University. H. Alon and D. Naveh thank the Israeli Ministry of Economy for funding within the "KAMIN" program under contract 50245. The authors thank the following for their assistance in the fabrication and characterization of samples: V. Artel, H. Aviv, O. Cohen, R. Cohen, Y. Erlich, M. Kirshner, M. Motiei, R. Popovtzer, S. Wissberg, and A. Zaban.